\documentstyle[11pt,newpasp,twoside,epsf]{article}
\reversemarginpar

\begin{document}
\title{The Structure of Cold Molecular Cloud Cores}
\author{D. Ward-Thompson \& D. J. Nutter}
\affil{Dept of Physics, Cardiff University, PO Box 913,
Cardiff, UK}
\author{J. M. Kirk}
\affil{Dept of Astronomy, University of Illinois,
Urbana, IL61801, USA}
\author{P. Andr\'e}
\affil{CEA/DSM/DAPNIA, Service d'Astrophysique, Saclay, France}

\begin{abstract}
A brief summary is presented of our current knowledge of the structure
of cold molecular cloud cores that do not contain protostars,
sometimes known as starless cores. The most centrally condensed
starless cores are known as pre-stellar cores.
These cores probably represent observationally
the initial conditions for protostellar collapse that must be input into
all models of star formation. The current debate over the nature of
core density profiles is summarised. A cautionary note is sounded over
the use of such profiles to ascertain the equilibrium status of cores.
The magnetic field structure of pre-stellar cores is also
briefly discussed.
\end{abstract}

\section{Introduction}

Low-mass (0.2--2.0M$_\odot$) protostar 
formation occurs in cold, dense cores within molecular clouds. However,
one of the major unknown factors in theories of star formation is a 
detailed observational determination of
the initial conditions of the collapse phase that forms a protostar.
There are many different theories of star formation, all
predicting different outcomes. Many of the differences
can be accredited to the fact that the
theories start from different
initial collapse conditions. The initial conditions
of collapse are crucially important to
all models of star formation (for a review, see Andr\'e, Ward-Thompson 
\& Barsony 2000).

Many studies of cold cores have been carried out to attempt to determine
the initial conditions observationally.
Molecular line surveys of dense cores by Myers and co-workers
identified a significant number of 
isolated cores (e.g. Myers \& Benson 1983;
Benson \& Myers 1989).
Comparison of these surveys 
with the IRAS point source catalogue
detected a class of core that had no associated infrared 
source. The lack of an embedded source led to the 
classification of these as `starless' cores (Beichman et al. 1986).

\begin{figure}
\includegraphics{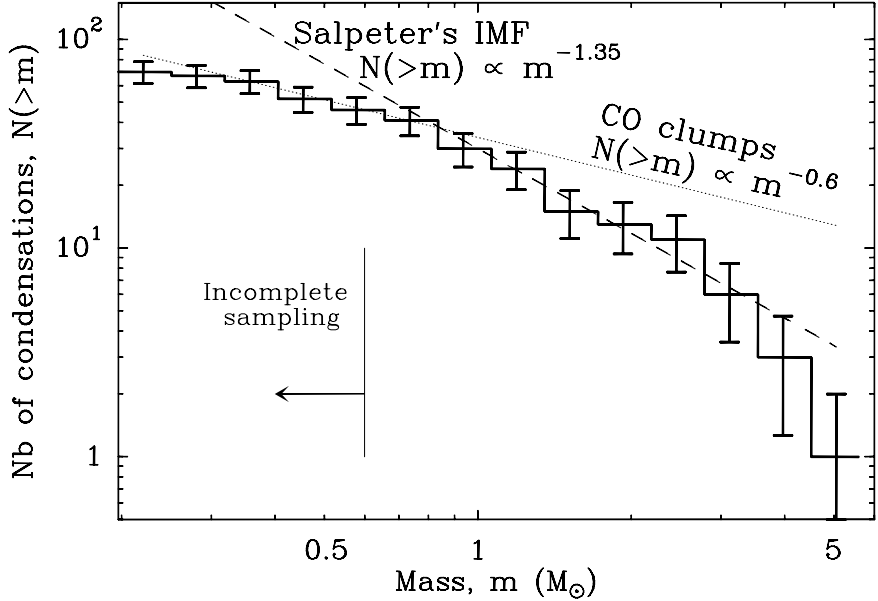}
\vspace*{8cm}
\caption{The clump mass distribution of the Orion B molecular cloud (from
Motte et al. 2001). Note that the distribution can be fitted by two 
power-laws, in a manner reminiscent of the stellar IMF. This appears
to indicate that the masses of stars are determined strongly by the masses
of the pre-stellar cores from which they form.}
\end{figure}

\begin{figure}
\includegraphics{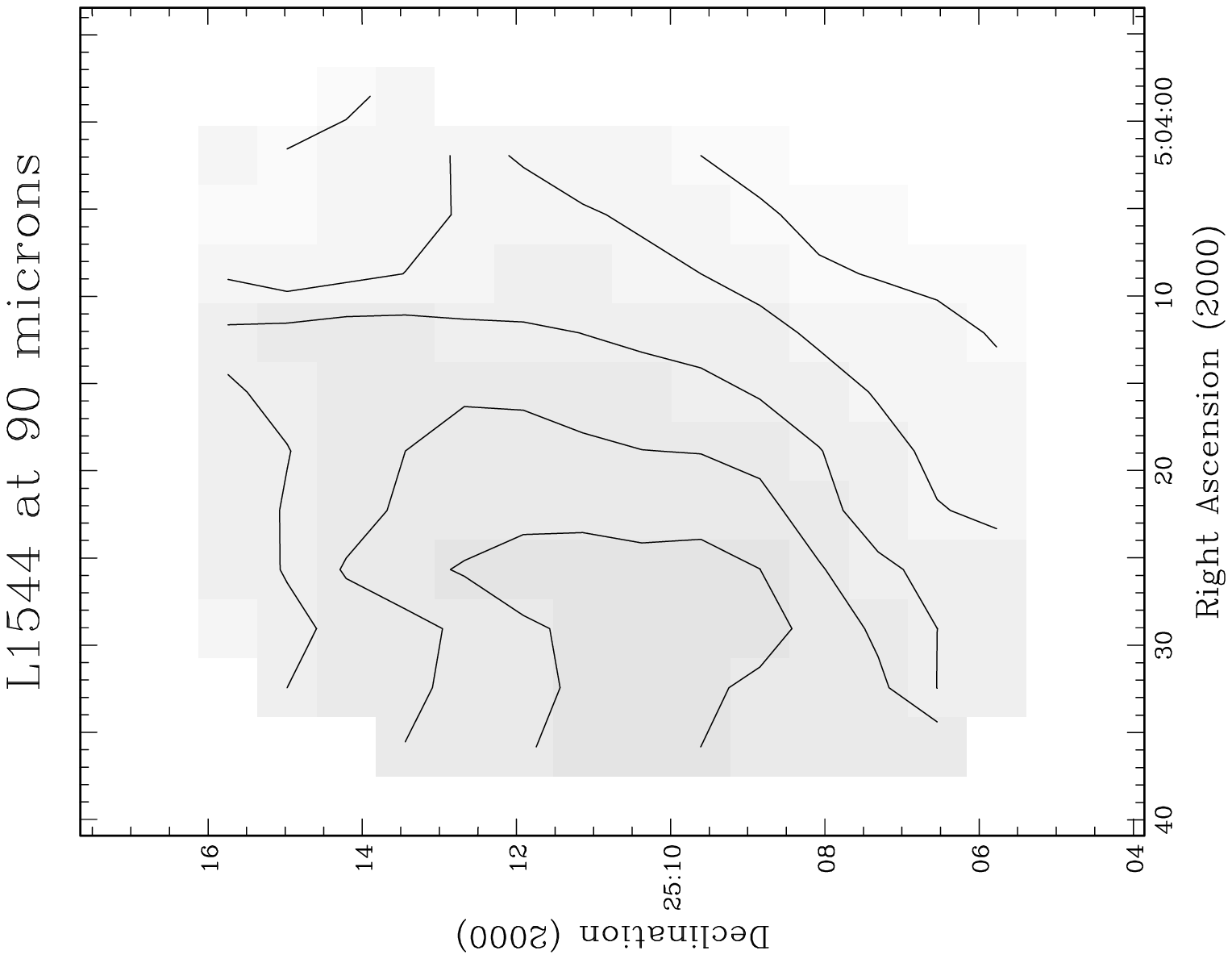}
\includegraphics{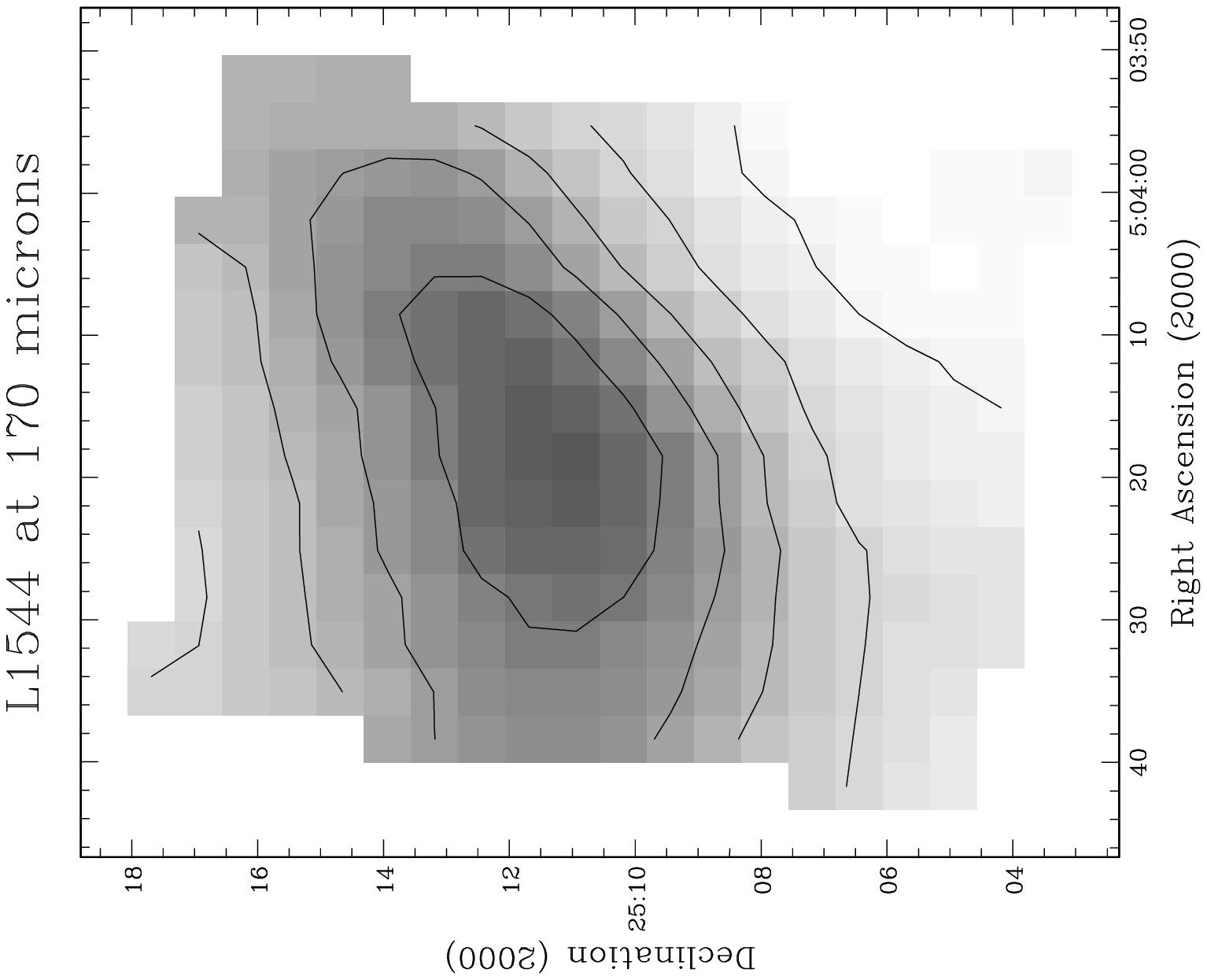}
\includegraphics{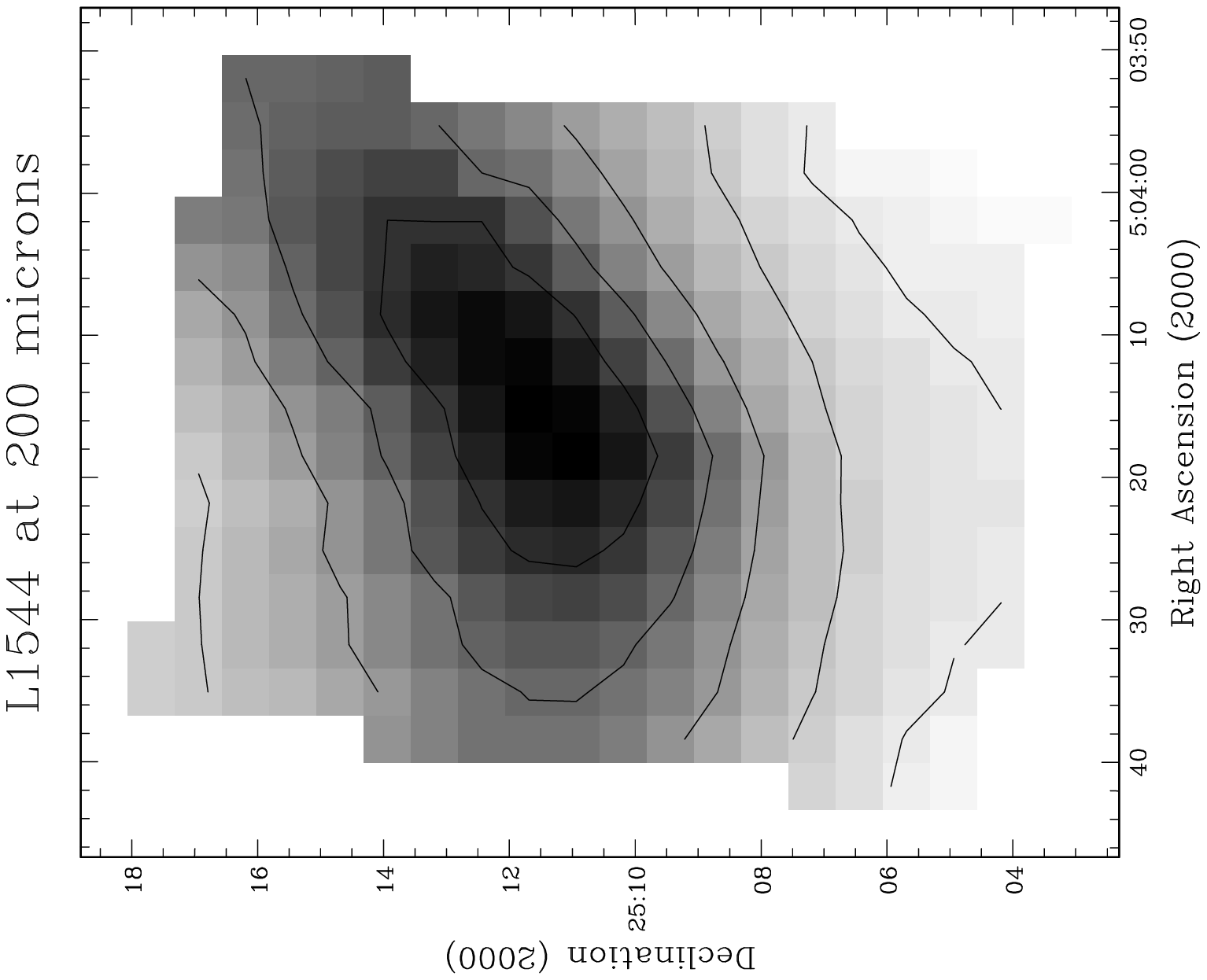}
\includegraphics{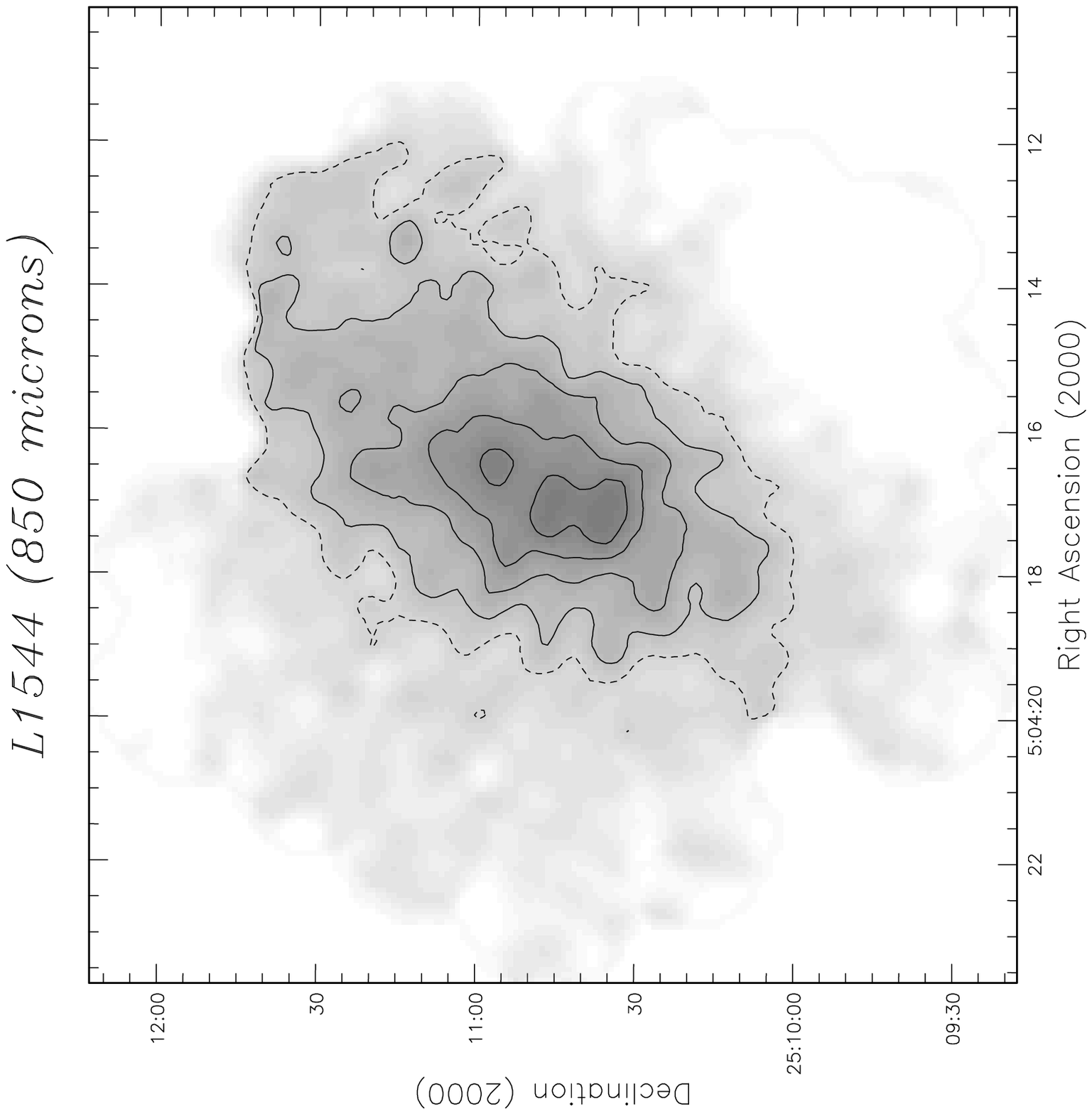}
\vspace*{18cm}
\caption{Images of the L1544 molecular cloud core at 90, 170 \& 200$\mu$m
as seen by ISO, and at 850$\mu$m as seen by SCUBA on the JCMT. 
The angular resolution
of the 850$\mu$m data is higher than that at the other wavelengths.
Note that the core is seen most clearly at the longer IR 
wavelengths, and that it is clearly non-circular.}
\end{figure}

Ward-Thompson et al. (1994)
coined the term pre-protosellar
cores to refer to those cores without embedded stars
that appear to be sufficiently centrally
condensed to be about to form stars. This term has subsequently
been shortened to `pre-stellar cores' for brevity.
Pre-stellar cores are clearly a key stage on the road
to star formation.

Some recent observations
have even indicated that the Initial Mass Function (IMF)
of stars may be determined at the pre-stellar core stage
(Motte, Andr\'e \& Neri 1998; Motte et al. 2001). Figure 1 shows
a plot of the mass distribution of the pre-stellar cores in
the Orion B molecular cloud region (from Motte et al. 2001).
The form of the distribution mimics the stellar IMF, apparently
indicating that the masses of stars are actually determined very
early in the star formation process, at the pre-stellar core
stage. If this proves to be correct, then in order to understand
the cause of the IMF, we must first understand the physics of
pre-stellar cores.

\section{Core morphologies}

Figure 2 shows a series of images of the L1544 pre-stellar core at
different far-IR and submm wavelengths. The core is barely visible at
90$\mu$m, but can be seen clearly
at longer IR wavelengths. In the submm,
the core appears somewhat fainter once again. This all indicates that
the core is very cold (see below). 

The morphology of L1544 can be seen from Figure 2 to be non-circular.
In the higher resolution 850$\mu$m data it can even be seen to appear
to break up into sub-structure. Nevertheless, many star formation
collapse models still start from spherically symmetric initial
conditions. However, some recent studies at least use elliptical
averaging to characterise the cores (see below).

\begin{figure}
\includegraphics{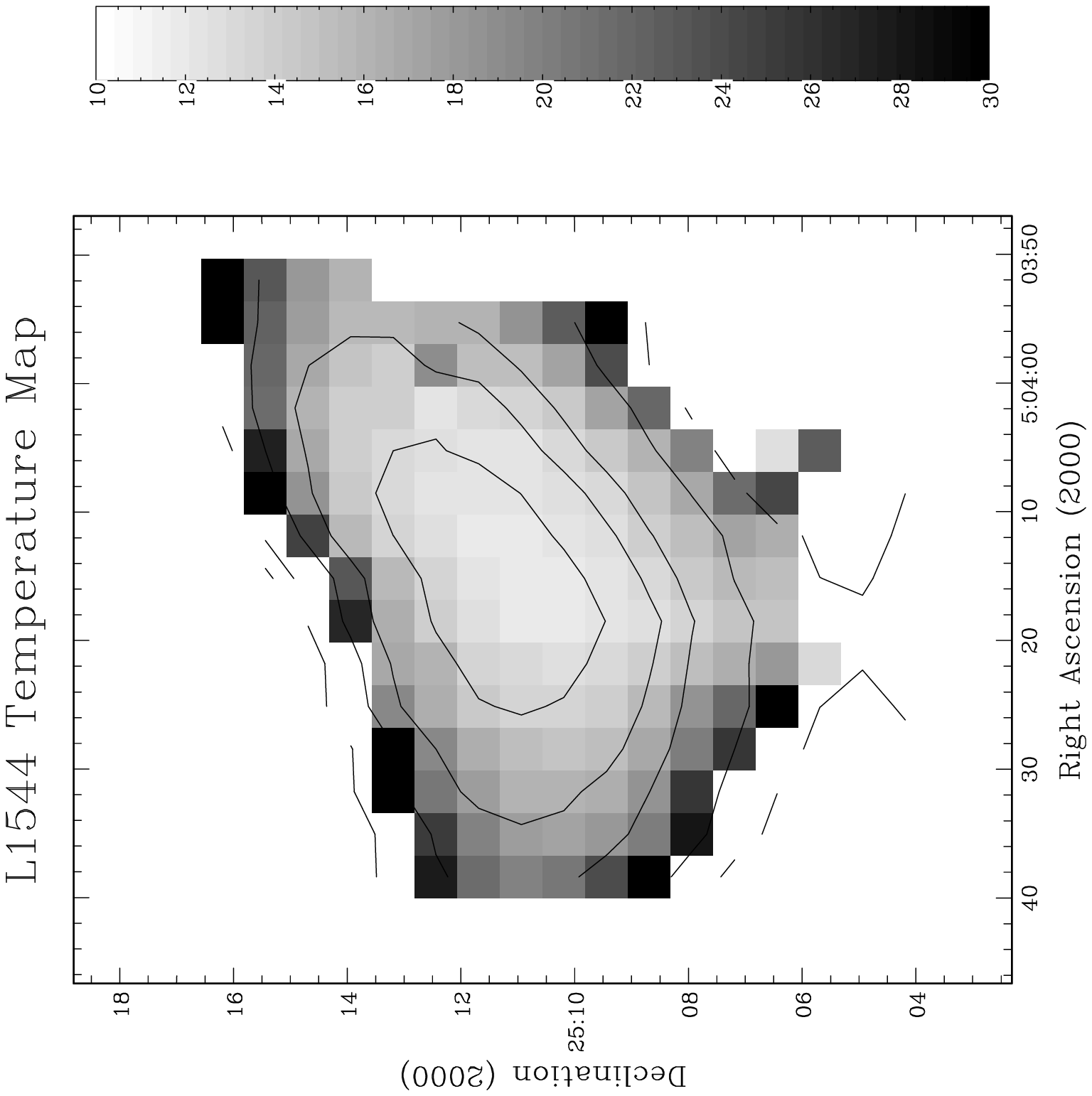}
\includegraphics{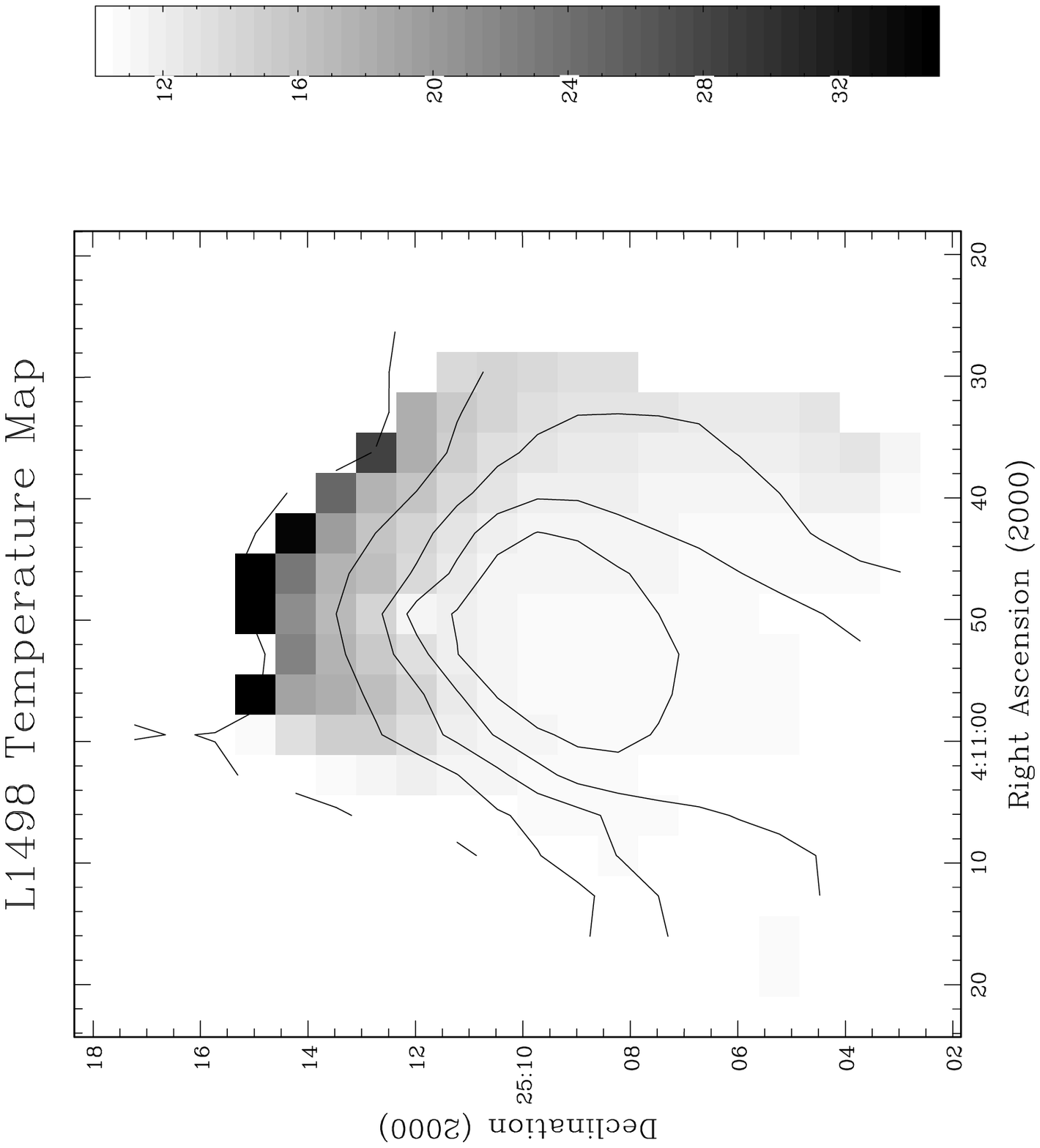}
\vspace*{8cm}
\caption{Colour temperature maps
of two pre-stellar cores, L1544 and L1498, constructed using the ratio
of the 200 and 170$\mu$m data.
Note that both cores appear warmer at the edge than at the centre.
This is consistent with cores that are externally heated, and have not
yet formed protostars in their centres.}
\end{figure}

\section{Temperature structure}

We can estimate the colour temperature variation across each
core (after subtraction of the background level) by ratio-ing the 
images at 170 and 200~$\mu$m. This ratio can be converted into
a colour temperature using the assumption of optically thin 
grey-body emission (see Ward-Thompson et al. 2002 for further
discussion), namely:

\[ (F_{\nu_1}/F_{\nu_2}) = 
\frac{\nu_{1}^{3+\beta}(e^{[h\nu_{2}/kT]} - 1)}
{\nu_{2}^{3+\beta}(e^{[h\nu_{1}/kT]} - 1)}, \]

\noindent
where the frequencies $\nu_{1}$ and $\nu_{2}$ correspond to
wavelengths of 200 and 170~$\mu$m respectively, and $F_{\nu_1}$ and
$F_{\nu_2}$ are the flux densities at each of these frequencies.
T is the dust temperature, $\beta$ is the dust emissivity index,
and $h$ and $k$ are the Planck and Boltzmann constants respectively.
This is one of the simplest sets of assumptions that can be made, as
it assumes that all of the dust in a 
given pixel is at a single temperature, T, and that it is
exactly the same dust which is emitting at both wavelengths.
We used $\beta$=2 to fit the data, as we
found this to be the typical value for these cores 
(Ward-Thompson et al. 2002).
Using these assumptions we constructed a series of colour temperature
maps.

Figure 3 shows the colour temperature maps we constructed for two cores,
L1544 and L1498. In both cases we see a temperature structure that features
a gradient which is cooler at the centre than the edge. This tends to
indicate that these cores show the behaviour
one would qualitatively expect for externally heated cores, and they
do not have a central protostar yet, as was suspected.
Hence they are confirmed as pre-stellar in nature. 

\begin{figure}
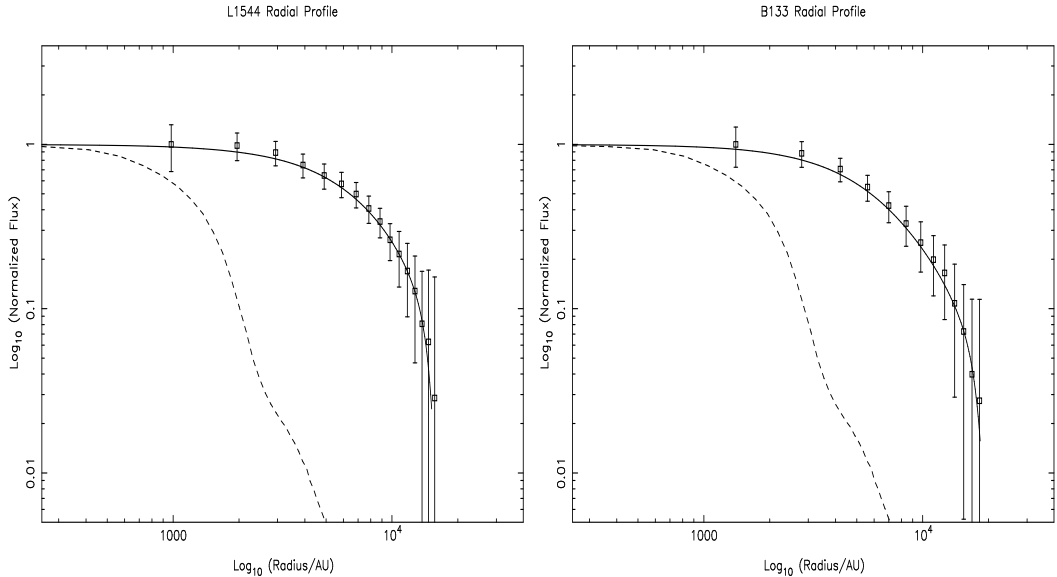

\includegraphics{wardthompson_d_fig4a.eps}
\includegraphics{wardthompson_d_fig4b.eps}
\vspace*{8cm}
\caption{Normalised logarithmic
radial flux density profiles of pre-stellar cores L1544 and B133
(elliptically averaged).
The profile is flat in the centre and steepens towards the edge in 
each case. This is typical of pre-stellar cores and has been modelled
in several ways.}
\end{figure}

\section{Density structure}

Ward-Thompson et al. (1994) were the first to find 
that pre-stellar cores all appear to follow a form of density profile 
that is relatively flat in the centre and steeper towards the edge. 
This is now seen to be a characteristic property of pre-stellar cores.
This profile cannot be explained as simply the result of the temperature
profile discussed above, as has been claimed (the observed submm flux
density profile is a multiple of temperature and density), since the
same shape of profile is seen in mid-IR
and near-IR absorption surveys (Bacmann et al., 2000; Alves et al., 2001).

%\begin{figure}
%\special{psfile=wardthompson_d_fig5.eps hoffset=-30 voffset=-400 hscale=80
%vscale=80 angle=0}
%\vspace*{8cm}
%\caption{850-$\mu$m image of L183 with polarised intensity vectors 
%superposed (taken with SCUBAPOL). 
%The vectors have been rotated by 90$^\circ$ to indicate
%the magnetic field direction. Note that the field is not aligned 
%with either the long axis or the short axis of the core.}
%\end{figure}

\begin{figure}
\includegraphics{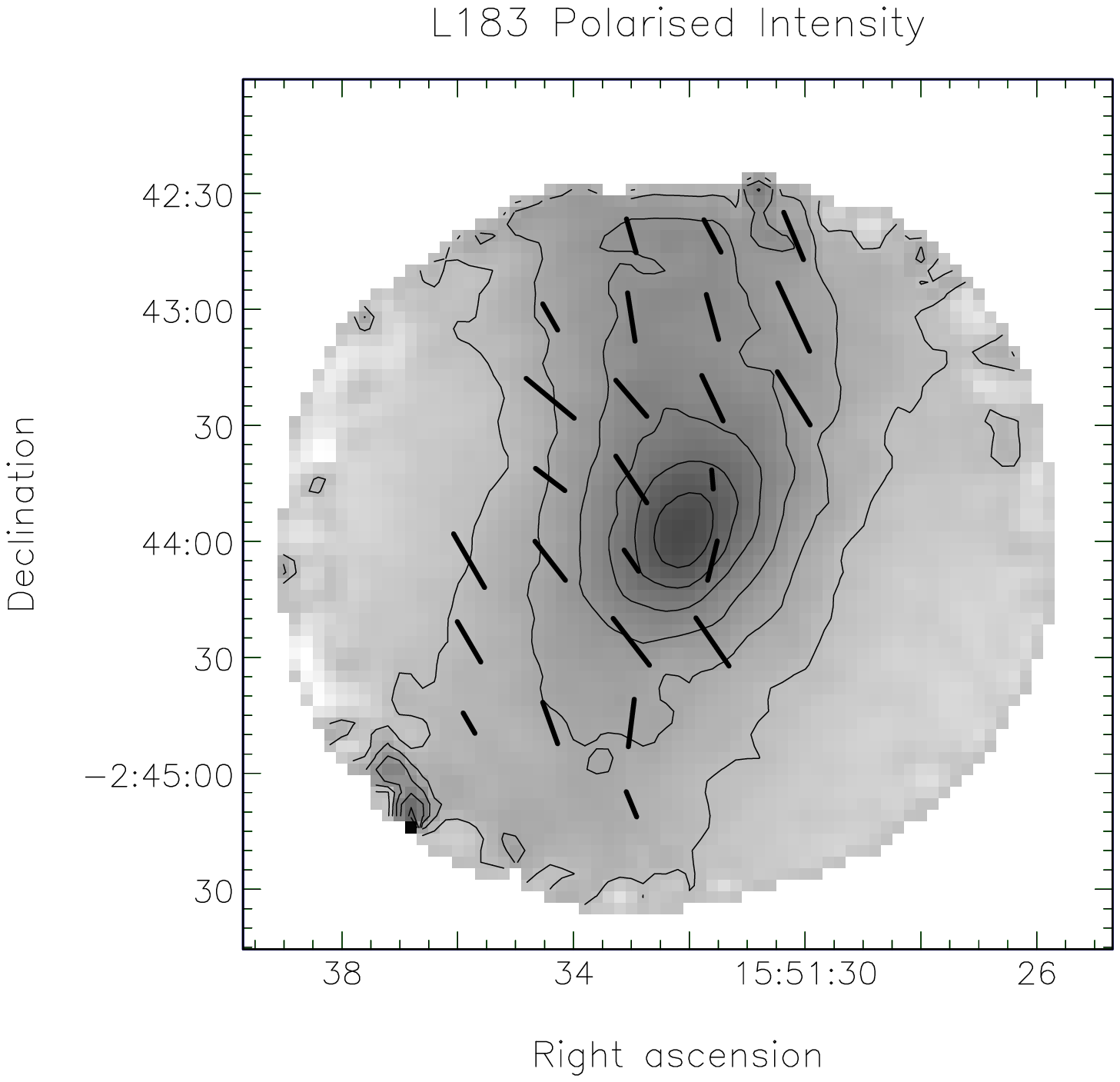}
\vspace*{8cm}
\caption{850-$\mu$m image of L183 with polarised intensity vectors
superposed taken with SCUBAPOL (from Crutcher et al., 2003).
The vectors have been rotated by 90$^\circ$ to indicate
the magnetic field direction. Note that the field is not aligned
with either the long axis or the short axis of the core.}
\end{figure}

Figure 4 shows the radial flux density profiles of two pre-stellar cores,
L1544 and B133, azimuthally averaged in elliptical bins to account for
the non-circularity of the cores.
The profiles can be seen to be flat in their
centres and steeper towards their edges
(for further discussion see: Kirk 2002; Kirk et al. 2003).
There have been several attempts to
characterise this form of profile, in terms of two power-laws 
(Ward-Thompson et al. 1994), multiple power-laws (Henriksen et al. 1997),
Plummer-type profiles (Whitworth \& Ward-Thompson 2001), or Bonnor-Ebert
profiles (Alves et al. 2001). 

The latter interpretation of density structure in terms of Bonnor-Ebert
profiles is interesting, because it might imply that the cores are in
some form of pressure equilibrium. However, this result may be misleading.
Some recent work has shown that even highly non-equilibrium cores can
demonstrate a Bonnor-Ebert form of profile 
(e.g. Ballesteros-Paredes et al. 2003).

Furthermore, it can be difficult to differentiate between forms of profile
based on a two-dimensional representation such as an astronomical image.
Recent work by Harvey et al. (2003a) has shown that for some data-sets
the profiles can often be fitted by many different forms of profile. 
This would tend to suggest that interpreting a particular
profile in terms of a pressure-balanced equilibrium may be over-interpreting
a limited amount of data.

Nonetheless, even if there may be ambiguity over exactly how to
model the density profiles, some forms of profile can be ruled out.
For instance both the singular isothermal sphere and the logotropic
non-isothermal sphere models can be ruled out (Bacmann et al., 2000).
The absolute value of the density at the centre and at the edge
can also provide information to help differentiate between models.

For example, in some cases a purely thermal
Bonnor-Ebert equilibrium model can be ruled because
the central temperature predicted by the model is much higher than
the observed temperature
(e.g. Andr\'e et al., 2003; Harvey et al., 2003b).
This could either be indicating that the cores are already collapsing
or that there is some additional form of support that is operating,
such as the support of a magnetic field.

\section{Magnetic fields}

The orientation of the magnetic field 
(in the plane of the sky) in pre-stellar cores can be ascertained
from submillimetre polarisation measurements of the cores.
Figure 5 shows a polarised intensity map of the L183 pre-stellar core
taken using SCUBA with its polarimeter SCUBAPOL (from Crutcher
et al., 2003). The vectors in
Figure 5 have been rotated by 90$^\circ$ to indicate the B-field
direction that can be inferred from the measured polarisation.

It can be seen from Figure 5 that the B-field is not aligned with
the short axis of the core, as would be predicted from quasi-static
models in which collapse occurs preferentially along the field lines.
Nonetheless the field appears moderately uniform across the core,
apparently ruling out highly turbulent motions tangling the field lines.

A detailed study of the dispersion of the vector orientations
(Crutcher et al., 2003) in this core concluded that the observations are 
consistent with magnetically-supported models of star formation in a weakly
turbulent medium. A comparison with turbulent magneto-hydrodynamic
simulations (e.g. Ostriker et al., 1999)
leads to similar conclusions (for further discussion
see Ward-Thompson et al., 2000). 
Hence to model the evolution
of these cores it appears we need to consider both the magnetic field and
turbulence.

\section{Conclusions}

We have given a brief summary of the current state of knowledge
of the structure of cold molecular cloud cores, in which solar-mass
stars are believed to form. We have concentrated on the properties of
pre-stellar cores -- those believed to be closest to protostellar
collapse. The main points can be summarised as follows:

\begin{itemize}

\item
Pre-stellar core morphologies are generally not spherically symmetric.

\item
Typical core temperatures are around 10K.

\item
Cores show a tendency to be cooler in their centres than at their edges.

\item
Core density profiles are flatter at their centres than at their edges.

\item
Core density profiles can typically be fitted by several
different analytic profiles.

\item
Core density profiles alone cannot be used to determine the equilibrium
status of pre-stellar cores.

\item
Core magnetic fields seem to be relatively uniform, but aligned with
neither the core major nor minor axes.

\end{itemize}

The latter point could be indicating that both turbulence and magnetic
fields play a role in the evolution of pre-stellar cores.
We support the continued study of pre-stellar cores, as they
represent our best opportunity for observing the
initial conditions for protostellar collapse.

\end{document}